# An FPGA-Based Semi-Automated Traffic Control System Using Verilog HDL

*Anik Mallik, Sanjoy Kundu, Md. Ashikur Rahman*
Department of Electronics & Communication Engineering, Khulna University of Engineering & Technology
E-mail: anikmallik@yahoo.com, sanjoykundu.ece@gmail.com, ashik_ece_kuet@yahoo.com

ABSTRACT
*Traffic Congestion is one of the severe problems in heavily populated countries like Bangladesh where Automated Traffic Control System needs to be implemented. An FPGA-based Semi-automated system is introduced in this paper including a completely new feature "Safe State" to avoid sudden unwanted collision. Here we used sequential encoding which has made the program much simpler and so that easy to control and modify. The experimental result showed the automated change in traffic lights according to the specified timing sequences which would be able to conduct maximum possible transitions of vehicles occur at different directions simultaneously without facing any accident.*

## 1. INTRODUCTION

Traffic Congestion problem is a common misery to the busy lives of the heavily populated countries like Bangladesh. Weak urban planning and Traffic Control System are mainly responsible for this problem. To form methods which will control the Congestion of traffic in the roads in urban and sub-urban areas remained the most challenging issue for the engineers [1]. So, research on this issue continues and various methods have been proposed by the researchers. The classical methods had mainly two drawbacks- one is, due to lack of perfect adjustments in the signal-timings, vehicles had to wait for a long time without any reason [2], and the other is, there was no such way in which ambulances, fire brigades, police vehicles and other emergency vehicles could pass with high level of priority [3]. Some of the recent methods used Fuzzy Logic to implement an intelligent system which would monitor and control the Traffic system [4]. These methods could set time adjustment and lead the cars with the help of arrival parameters through camera type sensors and image processing or electromagnetic sensors [5]. Fuzzy Logic is of a great use for systems whose appropriate mathematical modeling is not possible or very expensive to implement. However, the system may suffer from noticeable error if the input-output relation & the set of rules are weakly defined. So, there is a need to design a system, simple to operate, using FPGA based, cost effective technology which can eradicate these problems and can provide much greater speed.

Recently a Traffic Control System has been developed using Mealy State Machines [6]. But, the system grown in that research, has limited Emergency States for passing vehicles on the roads, which had different switches for different roads and had no "Safe State" (i.e. all roads having alert signal to avoid sudden unwanted collisions) between an emergency and normal or ordinary state. The system also suffered from inadequate Green signals for various directions and signal for human-passing on Zebra-crossings. In our proposed design, an input of 3 binary bits has been provided which is used for defining the states, including a new idea of "Safe State". Emergency states are changed manually and too easy to handle. The major or minor roads can be defined manually and the system can run that way. Timings are specified here based on our assumptions. This can also be changed by a statistical approach depending upon the necessity of the roads in practical. We also have included signals for Zebra-crossing into the system and this design allows the maximum possible transitions of vehicles on different directions without facing any accident along with the basic features found in any control system, which can be fabricated into one single chip.

The system has been written in Verilog HDL [7] & designed, tested and evaluated using the ISE 6.0 tool of Xilinx and VeriloggerPro 6.5 [8]. Then the design has been implemented on Xilinx Spartan-2 FPGA (XC2S150). In the following sections, the methodology, the implementation technique of this system, designed circuit schematic and the result of the experiment will be discussed in brief.

## 2. METHODOLOGY
### 2.1 Mealy State Machine & State Machine Style
State machines can be classified into two categories: Mealy machines and Moore machines. Basically, the process of generation of the output differentiates between Mealy and Moore machines. In a Mealy machine, the outputs are generally a function of both the state and the inputs [9].

The State Machine Style can be of various forms. One of these forms combines the next-state logic and the state register. This style is good to use since the next-state logic and state register here are related quite inseparably. If the state machine has many inputs which change frequently, this style is efficient enough to use [9].

### 2.2 Sequential Encoding
State encoding has a great effect on the size and performance of a circuit, and also influences the amount of electronic glitches produced by a circuit. This can also be of several kinds: Sequential Encoding, Gray Encoding, One-Hot or One-Cold Encoding etc.

In Sequential Encoding, state comes one after another according to the sequence of bit pattern and makes the system simpler and efficient. But glitches may occur when we deal with a circuit with huge dimensions. In case of Gray Encoding, the simplicity in program code along with the circuitry is missed as it does not change the bit pattern sequentially and provides speed little bit slower. One-Hot or One-Cold Encoding is very efficient when we need to consider faster operation but in this case the area of circuitry becomes huge [9].

In this design, we have used Sequential Encoding as our concern was to simplify the operation. The proposed circuitry is not huge, so this scheme gave us the optimized performance.

### 2.3 Field Programmable Gate Array (FPGA)
A Field Programmable Gate Array (FPGA) is an integrated circuit and its configuration is generally defined using a Hardware Description Language (HDL), and in this case, it is same as an Application Specific Integrated Circuit (ASIC). This name 'FPGA' is given as it is structured like a gate array configuration of an ASIC [10]. This regular structure is interconnected which is under the designers' extreme control, which means that the user can design, modify or make any change to his circuit easily [11]. FPGAs can be used to implement any logical function that can be performed by an ASIC, and it should also be noted that FPGA design is more cost-effective than that of ASIC [12]. They have lots of advantages over microcontrollers, such as greater speed, number of I/O ports and performance [13].

### 3. SYSTEM IMPLEMENTATION
#### 3.1 States
The simplest form of traffic light is introduced with a single or a pair of colored aspects that warns any vehicle of the shared right of way of a possible conflict or danger [14]. But, in case of a complex form, the scenario is a bit different.

#### 3.1.1 Traditional States
Now, we have to imagine a junction of four roads. In this complex state, each road will have six lights (Red, Yellow, Green for going straight, Green for turning Left and Green for turning Right and Green for Zebra-crossing) which are illustrated in **Fig. 1**. The timing sequence is shown in the **Table 1**.

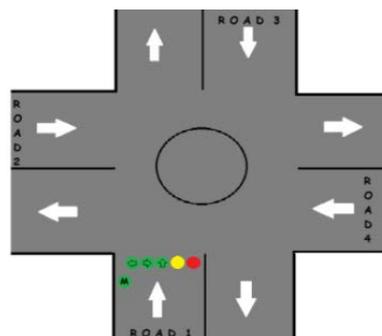

**Fig. 1:** A Four way road each having six lights

**Table 1**: Timing Sequence of Traditional States (Only Road 1 is shown here)

| Timing Sequence | R1 | Y1 | G1 ↓ | G1 → | G1 → | M1 | HEX (for all the four roads) |
|---|---|---|---|---|---|---|---|
| 60 sec | 0 | 0 | 1 | 1 | 0 | 0 | 3218A6 |
| 15 sec | 0 | 1 | 0 | 0 | 0 | 0 | 410820 |
| 60 sec | 1 | 0 | 0 | 1 | 1 | 0 | 98C862 |
| 15 sec | 1 | 0 | 0 | 0 | 0 | 0 | 810420 |
| 60 sec | 1 | 0 | 0 | 0 | 1 | 0 | 8A6321 |
| 15 sec | 1 | 0 | 0 | 0 | 0 | 0 | 820410 |
| 60 sec | 1 | 0 | 0 | 0 | 0 | 1 | 86298C |
| 15 sec | 0 | 1 | 0 | 0 | 0 | 0 | 420810 |

#### 3.1.2 Emergency States
In some special cases, like emergency states, the controlling would be different from the traditional one. For example, we can easily open Road 1 for having an emergency in this way, and can block the others which are shown in the **Table 2**.

**Table 2:** Timing Sequence of Emergency-1 State
(Only Road 1 is shown here)

| Timing Sequence | R1 | Y1 | G1 ↓ | G1 → | G1 ← | M1 | HEX (for all the four roads) |
|---|---|---|---|---|---|---|---|
| 15 sec | 0 | 1 | 0 | 0 | 0 | 0 | 410410 |
| ≥ 60 sec | 0 | 0 | 1 | 1 | 1 | 0 | 3A0822 |

*3.1.3 Safe State*

The Safe State is needed to occur between the transitions of traditional state and emergency one. This is shown in **Table 3**.

**Table 3:** Timing Sequence for "Safe State"
(Only Road 1 is shown here)

| Timing Sequence | R1 | Y1 | G1 ↓ | G1 → | G1 ← | M1 | HEX (for all the four roads) |
|---|---|---|---|---|---|---|---|
| ≥ 15 sec | 0 | 1 | 0 | 0 | 0 | 0 | 410410 |

**3.2 System Flow Chart**

The proposed system can be expressed through the Flow Chart in **Fig. 2**.

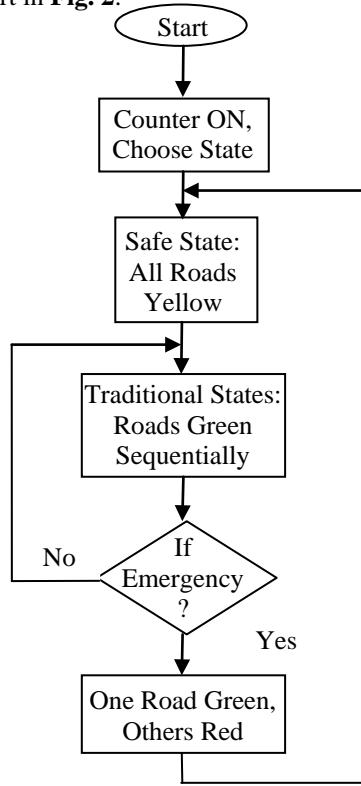

**Fig. 2:** Flow Chart for our prescribed System

**3.3 Circuit Schematic**

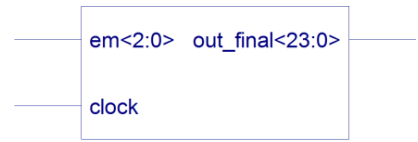

**Fig. 3:** Block Diagram of FPGA-Based Semi-Automated Traffic Control System

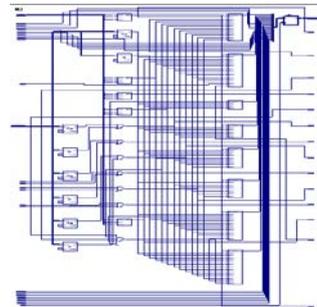

**Fig. 4:** RTL Schematic of one of the system blocks

**4. RESULTS**

We used the Verilog as HDL to design this system and then verified it in VeriloggerPro 6.5 software. The timing diagram achieved from Testbencher showed that the traffic lights were being ON and OFF automatically according to the timing sequences. Then we tested it in the Xilinx FPGA where it also gave the correct output. The timing diagrams are provided in the next sub-section. The 24 bits of outputs are converted here into 6-digits Hexadecimal numbers.

**4.1 Timing Diagrams**
*4.1.1 Traditional State*

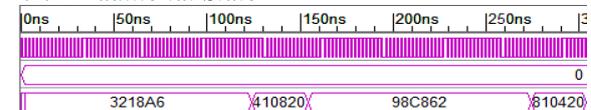

**Fig. 7:** Timing Diagram of Traditional State Output (from 0ns to 300ns)

*4.1.2 Emergency State*

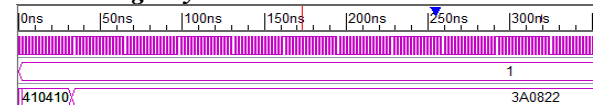

**Fig. 8:** Timing Diagram of Emergency-1 State Output (from 0ns to 350ns)

*4.1.3 "Safe State"*

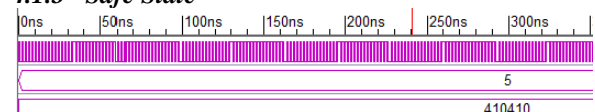

**Fig. 9:** Timing Diagram of "Safe State" Output (from 0ns to 350ns)

## 5. CONCLUSION

We have presented an FPGA-Based Semi-Automated Traffic Control System, where we used Verilog as the HDL. We have named this design as "Semi-Automated" because we could not make the emergency-state input system fully automated. We used the Mealy State Machine and the Sequential Encoding made the program simplest. This design allows the maximum possible transportation of vehicles in different directions simultaneously. One of the salient features of this design is any addition of different lights for different purposes can take place easily just by adding bits to the output ports. Again, we have left two more states in this design which can be used for other necessary purposes. The whole design needs only two inputs- one is Clock and the other is for different states, which can be controlled fluently. The most effective feature of this design is the use of 15sec "Safe State" which is able to avoid the sudden unwanted collision between vehicles in the transitions of states easily. Any kind of step for the improvement of this system can easily be taken for this design because the logic is made simpler than ever.

## 6. FUTURE WORK

We have a plan to include a feature to this system which will be able to count the number of the vehicles in a road. This can be accomplished using a number of Heat-sensors or IR-sensors placed in different positions in a road. Each road will be timed automatically according to the number of vehicles.

## 7. ACKNOWLEDGEMENTS

We want to acknowledge gratefully Mr. Robin Sarker (B.Sc. Engr.), Mr. Shumit Saha (Lecturer) and Dr. A. B. M. Aowlad Hossain (Asst. Professor) from Dept. of Electronics & Communication Engineering, Khulna University of Engineering & Technology, for their valuable suggestions & also for providing the facilities to implement the design on FPGA.